\documentclass[11pt,twoside]{article}

\usepackage{asp2006}
\usepackage{lscape}
\usepackage{graphicx}

\begin{document}
\title{Do AGN suppress star formation in early-type galaxies?}   
\author{Kevin Schawinski}   
\affil{Einstein Fellow}
\affil{Department of Physics, Yale University, New Haven, CT 06511, U.S.A.}    
\affil{Yale Center for Astronomy and Astrophysics, Yale University, P.O. Box 208121, New Haven, CT 06520, U.S.A.}
\begin{abstract} 
The observation that AGN host galaxies preferentially inhabit the ``green valley" between the blue cloud and the red sequence has significant consequences for our understanding of the co-evolution of galaxies and black holes via accretion events. I discuss the interpretation of green valley AGN host galaxy colours with particular focus on early-type galaxies.
\end{abstract}

\def\Chandra{\textit{Chandra}}
\def\XMM{\textit{XMM-Newton}}
\def\Swift{\textit{Swift}}

\def\OI{[\mbox{O\,{\sc i}}]~$\lambda 6300$}
\def\OIII{[\mbox{O\,{\sc iii}}]~$\lambda 5007$}
\def\SII{[\mbox{S\,{\sc ii}}]~$\lambda \lambda 6717,6731$}
\def\NII{[\mbox{N\,{\sc ii}}]~$\lambda 6584$}

\def\Ha{{H$\alpha$}}
\def\Hb{{H$\beta$}}

\def\NIIHa{[\mbox{N\,{\sc ii}}]/H$\alpha$}
\def\SIIHa{[\mbox{S\,{\sc ii}}]/H$\alpha$}
\def\OIHa{[\mbox{O\,{\sc i}}]/H$\alpha$}
\def\OIIIHb{[\mbox{O\,{\sc iii}}]/H$\beta$}

\def\Ebmv{E($B-V$)}
\def\LOIII{$L[\mbox{O\,{\sc iii}}]$}
\def\Ledd{${L/L_{\rm Edd}}$}
\def\LOIIIs4{$L[\mbox{O\,{\sc iii}}]$/$\sigma^4$}
\def\LOIIIMbh{$L[\mbox{O\,{\sc iii}}]$/$M_{\rm BH}$}
\def\Mbh{$M_{\rm BH}$}
\def\Msigma{$M_{\rm BH} - \sigma$}
\def\Ms{$M_{\rm *}$}
\def\Msun{$M_{\odot}$}
\def\Msunyr{$M_{\odot}yr^{-1}$}

\def\ergs{$~\rm ergs^{-1}$}
\def\kms{$~\rm kms^{-1}$}



\section{Introduction}
Recent observations of the host galaxies of AGN both in the nearby universe and at redshifts out to $z \sim 1$ have shown that the restframe optical colours of the host galaxies peak at intermediate colours between the blue cloud of star-forming galaxies and the red sequence of passively evolving galaxies \citep[e.g.][]{2007MNRAS.382.1415S, 2007ApJ...660L..11N, 2008ApJ...675.1025S}. This observation has led to the suggestion that these ``green valley'' AGN host galaxies represent a population in transition from the blue cloud to the red sequence that continuously build up the red sequence across a large range in cosmic time \citep[e.g.][]{2007ApJ...665..265F, 2007MNRAS.382.1415S, 2008ApJ...681..931B}. However, the interpretation of green or intermediate host galaxy colours is not necessarily unique.

\section{What does the green valley really mean?}
In general, green or intermediate optical colours can arise from a number of scenarios. The three main scenarios are (1) the recent, rapid suppression of star formation; (2) the slow, gentle fading of star formation; and (3) an enhanced dust screen covering a regular blue cloud star-forming galaxy. Thus the observation that AGN host galaxies exhibit green optical colours is consistent with a scenario where a recent catastrophic event (such as the AGN phase) has shut down star formation; more details work on the host galaxy stellar populations are needed to determine which scenario has led AGN host galaxies to the green valley.

However, if the green colours of AGN host galaxies are due to a recent, rapid suppression of star formation, then the fact that the host galaxies are green leads to a time scale problem. Green colours in such a suppression scenario imply that the event that led to the suppression of star formation is already significantly in the past. The minimum amount of time elapsed since the suppression event in the case of instantaneous suppression is set by the lifetime of OB stars that dominate the ultraviolet-optical spectral energy distribution of young stellar populations. Thus, if there are no AGN host galaxies with blue host galaxy colours, then the AGN population detected in the green valley cannot be responsible for the suppression of star formation. 

A study of an unbiased, hard X-ray selected sample of local AGN host galaxies detected by the \textit{Swift} Burst Alert Telescope by \cite{2009ApJ...692L..19S} shows that the local high luminosity $L_{14-195~ \rm keV} > 10^{42.7}$\ergs\ AGN host galaxies, even when accounting for any contribution of blue AGN continuum in the nucleus, strongly cluster at intermediate colours. The distribution of these colours, assuming a scenario of rapid suppression of star formation, imply a minimum time delay of $\sim 100$ Myr from the suppression event to the emergence of a radiatively efficient Seyfert AGN phase detected in the hard X-ray band of \textit{Swift}.

\section{The star formation history of early-type AGN host galaxies}

A more detailed study of the stellar populations of AGN host galaxies has been performed by \cite{2007MNRAS.382.1415S}. Using both broadband photometry from the ultraviolet (from \textit{GALEX}), optical (SDSS) and near-infrared (2MASS) and stellar absorption indices measured from SDSS spectra, \cite{2007MNRAS.382.1415S} determined both the age and mass-fraction of the most recent episode of star formation in emission-line selected Seyfert AGN with early-type host galaxies. They found that early-type Seyfert AGN host galaxies typically experienced a substantial starburst that built typically 1--10\% of the total stellar mass between 300 Myr and over 1 Gyr in the past. None of the early-type AGN host galaxies are experiencing a co-eval episode of star formation. They are, on the other hand, part of an evolutionary sequence that begins with a starburst in the blue cloud and ends $\sim 1$ Gyr later on the red sequence. Early-type AGN host galaxies can thus be described as post-starburst objects.

Follow-up observations of the molecular gas content of the early-types  using the CO(1$\rightarrow$0) line with the 30m IRAM telescope \citep{2009ApJ...690.1672S} furthermore show that by the time of the Seyfert phase, the formerly substantial cold gas reservoirs ($\sim10^9$\Msun) present during the starburst phase has been destroyed. The destruction of the molecular gas reservoir takes place during the previous AGN+SF composite phase (the transition region on the BPT diagram) where a low-luminosity AGN and rapidly declining star formation are roughly comparable in their contribution to the budget of ionizing photons.  It would have to be this earlier, low-luminosity AGN phase that could potentially be responsible for the suppression of star formation by destroying the molecular gas reservoir, perhaps in a radiatively inefficient kinetic feedback mode.

\section{Summary}
The fact that AGN host galaxies preferentially exhibit green valley optical colours does not necessarily imply that they have experienced a recent, dramatic shutdown of star formation, and other interpretations of the broad-band colours are possible. In addition, green colours imply that the star formation has \textit{already} been shut down at least $\sim 100$ Myr in the past. A detailed analysis of the stellar populations of \textit{early-type} AGN host galaxies from the Sloan Digital Sky Survey shows that early-type AGN hosts are post-starburst galaxies with ages of 300 Myr to over 1 Gyr. They form part of an evolutionary sequence starting in the blue cloud and ending on the low-mass end of the red sequence. 

\acknowledgements 
Support for the work of K.S. was provided by NASA through Einstein Postdoctoral Fellowship grant number PF9-00069 issued by the Chandra X-ray Observatory Center, which is operated by the Smithsonian Astrophysical Observatory for and on behalf of NASA under contract NAS8-03060. K.S. gratefully acknowledges support from Yale University. 


\end{document}